\documentstyle[preprint,aps,amsfonts,epsf]{revtex}

\input{epsf.tex}

\newcommand{\s}{\sigma}
\newcommand{\I}{{\rm i}}
\renewcommand{\d}{{\rm d}}
\renewcommand{\a}{\alpha}

\newcommand{\dfrac}[2]{\displaystyle\frac{#1}{#2}}
\newcommand{\be}{\begin{equation}}
\newcommand{\ee}{\end{equation}}
\newcommand{\bea}{\begin{eqnarray}}
\newcommand{\eea}{\end{eqnarray}}
\newcommand{\ba}{\begin{array}}
\newcommand{\ea}{\end{array}}
\def\J#1#2#3#4{#1 {\bf #2}, #3 (#4)}

\def\ib{ib.}

\def\PR{Phys. Rev.}

\def\AJ{Astrophys. J.}

\def\MN{M.N.R.A.S.}
\def\BAN{Bull. Astron. Inst. Neth.}
\def\ARAA{Ann. Rev. Astron. Astroph.}

\def\LOB{Lick. Obs. Bull.}

\def\VA{Vistas in Astronomy}

\begin{document}
\draft
\title{The Rotation Curve and Mass--Distribution \\
in Highly Flattened Galaxies}

\author{N.~R.~Sibgatullin,\thanks{On leave of absence from the
Department of Hydrodynamics, Lomonosov Moscow State University,
Moscow 119899, Russia} A.~A.~Garc\'\i a, V.~S.~Manko}
\address{Departamento de F\'\i sica, Centro de Investigaci\'on y de
Estudios Avanzados del IPN,\\ A.P.~14-740, 07000 M\'exico D.F.,
Mexico}

\maketitle

\begin{abstract} A new method is developed which permits the
reconstruction of the surface--density distribution in the
galactic disk of {\it finite} radius from an {\it arbitrary}
smooth distribution of the angular velocity via two simple
quadratures. The existence of upper limits for disk's mass and
radius during the analytic continuation of rotation curves into
the hidden (non--radiating) part of the disk is demonstrated.
\end{abstract}

\bigskip

\hspace{4cm}PACS number(s): 98.50

\newpage


The rotation curves of spiral galaxies supply astronomers with the
direct information about distribution of the gravitational
potential in galactic disks. The measurements of H$_\a$, HI and CO
lines allowed to construct rotation curves for our and nearby
galaxies (for early reviews cf. Refs.~[1--3]). The advanced modern
techniques of observation and data processing permitted to extend
rotation curves at the distances far beyond the galactic center
and discover the flat rotation curves [3--7]. The first galactic
disk models were based on the concept of oblate shells (homoeoids)
between confocal spheroids [8--10]. Brandt \cite{Bra}, Brandt and
Belton \cite{BrB} tended the eccentricity of confocal spheroids to
zero, thus obtaining the disks covering the whole plane $z=0$.
They proposed a special rotation curve possessing Keplerian
asymptotics for M31 and NGC5055, and found numerically the
corresponding surface--density distribution. Another approach was
developed by Toomre \cite{Too} who considered a new class of
asymptotically Keplerian rotation curves and found, for infinite
disks, the corresponding surface--density distributions. The
problem of reconstruction of the surface density from the rotation
curve in the case of infinite disks containing a black hole at the
center was solved in our work \cite{SGM}. Mestel \cite{Mes}
pointed out that the highly flattened galaxies and nebulas should
be modeled by self--gravitating disks of {\it finite} radius $R$.
He discussed two limiting cases: a rigidly rotating disk and a
model with a constant velocity. In a different context some exact
models of self--gravitating disks with outer boundary were found
by Morgan and Morgan \cite{MM}.

In the present paper we propose a new method to the problem of
reconstruction of the surface--density distribution in a
self--gravitating disk of finite extension from an arbitrary
smooth distribution of the angular velocity, and it permits to
obtain the {\it general} solution of this problem in terms of two
simple quadratures. We shall also demonstrate that if a rotation
curve is analytically continued into the hidden (non--radiating)
part of the disk, then the disk's mass can achieve its maximal
value at a {\it finite} radius, and not necessarily at infinity.
Like it is done in the classical papers [10--13], we shall assume
an idealized model for the highly flattened galactic disks in
which the influence of viscosity, pressure, non--disk component
and all sort of non--circular motion of matter will be neglected.
We will not separate the gravitational fields of the dark
(non--radiating) and radiating matter, firstly, because they both
exert the same influence on rotation curves, and, secondly,
because no reliable mechanism for such separation is yet known
[17, 4, 18].


Our approach to the Newtonian self--gravitating disk problem is
based on a new representation for the gravitational potential of a
disk with the boundary radius $R$ which we take in the form
\be
\Phi(\rho,z)=\sqrt{\tilde\rho^2+\tilde z^2}\,\,
\tilde\Phi(\tilde\rho,\tilde z)\,, \ee where the auxiliary space
variables $\tilde\rho,\tilde z$ are related to the genuine
cylindrical coordinates $\rho,z$ via the formulas
\be
\tilde\rho=R^2\rho/r^2, \quad \tilde z=R^2z/r^2, \quad r^2\equiv
\rho^2+z^2\,, \ee and the function $\tilde\Phi$ is defined in the
following way:
\be
\tilde\Phi(\tilde\rho,\tilde
z)=\frac{1}{2\pi^2}\int\limits_0^{\pi} \d\theta
\int\limits_R^\infty \ln[(s-\tilde\rho\cos\theta)^2+\tilde
z^2]\,\a(s)\,\d s\,, \ee $\a(s)$ being a real function which has
the meaning of the density of sources distribution. Due to the
specific analytical properties of the logarithmic function which
turn out to be most advantageous for fulfilling the Laplace
equation in the region exterior to disk's sources, hereafter we
shall be able to relate the function $\a(s)$ to the surface
density $\s(\rho)$ of the disk and to the distribution of the
angular velocity $\omega(\rho)$ via simple integral formulas.

The potential $\Phi(\rho,z)$ which, as can be easily seen from
(1), is connected with $\tilde\Phi$ by Thomson's transform,
satisfies Laplace's equation everywhere except on the disk. On the
circle $z=0$, $\rho\le R$ the normal derivative
$\partial\Phi/\partial z$ experiences a jump. To show this, let us
tend $z$ to zero. Then, by virtue of (2), we have
\be
\lim_{z\to+0}\frac{\partial\Phi}{\partial z} =
\frac{\tilde\rho^3}{R^2}\lim_{\tilde z\to+0}
\frac{\partial\tilde\Phi}{\partial\tilde z}\,, \ee while from (3)
follows that
\be
\lim_{\tilde z\to+0}\frac{\partial\tilde\Phi}{\partial\tilde
z}=\lim_{\tilde z\to+0}\frac{1}{2\pi^2}
\int\limits_0^{\pi}\d\theta\int\limits_R^\infty\left(
\frac{-\I}{s-\tilde\rho\cos\theta-\I\tilde
z}+\frac{\I}{s-\tilde\rho\cos\theta+\I\tilde z}\right) \a(s)\,\d
s\,. \ee

Due to the Sokhotsky--Plemelj formula for the Cauchy-type
integrals, we obtain from (5)
\be
\left.\frac{\partial\tilde\Phi}{\partial\tilde z}\right|_{\tilde
z\to+0}= \left\{
\begin{array}{ll}0\,,&0\le\tilde\rho<R\\
\dfrac{1}{\pi}\int\limits_R^{\tilde\rho} \dfrac{\a(s)\d
s}{\sqrt{\tilde\rho^2-s^2}}\,,&\tilde\rho>R\,,\\
\end{array}\right. \ee or, returning to the coordinates $\rho,z$:
\be
\left.\frac{\partial\Phi}{\partial z}\right|_{z\to+0}= -2\pi G
\s(\rho)=\left\{
\begin{array}{ll} \dfrac{R^4}{\pi\rho^2}\int\limits_R^{R^2/\rho}
\dfrac{\a(s)\d s}{\sqrt{R^4-\rho^2s^2}} \,,&0\le\rho<R \\
0\,,&\rho>R\,,\\
\end{array}\right. \ee where $\s(\rho)$ is the surface density of
the disk.

The condition of balance of the gravitational and centrifugal
forces for zero--pressure matter of the disk at $z=0$, $\rho<R$,
i.e.,
\be
\left.\frac{\partial\Phi}{\partial \rho}\,\right|_{z\to0}
+\omega^2\rho =0\,, \ee can be represented in the form
\be
\frac{\partial}{\partial\tilde\rho}(\tilde\rho\,\tilde\Phi)=
\frac{\omega^2R^4}{\tilde\rho^3}\,, \ee whence it follows that
\be
\tilde\rho\frac{\partial\tilde\Phi}{\partial\tilde\rho} =
\frac{R^4}{\tilde\rho}\int\tilde\rho\left(
\frac{\omega^2}{\tilde\rho^3}\right)_{,\tilde\rho}\d\tilde\rho\,.
\ee

From (3) we obtain the form of the left--hand side of (10) in the
limit $z\to0$:
\be
\tilde\rho\frac{\partial\tilde\Phi}{\partial\tilde\rho} =
\frac{1}{\pi}\left(\int\limits_R^\infty \a(s)\,\d s-
\int\limits_{\tilde\rho}^\infty\frac{\a(s)s\,\d
s}{\sqrt{s^2-\tilde\rho^2}} \right)\,. \ee

In the subsequent formulas it is convenient to pass to the
non--dimensional variables
\be
t\equiv R^2/s^2\,, \quad x\equiv R^2/\tilde\rho^2=\rho^2/R^2\,.
\ee

In the new variables, formula (7) for the surface--density
distribution assumes the form
\be
\s(x)=-\frac{R}{4\pi^2 Gx}\int\limits_x^1\frac{\a(t)\,\d
t}{t\sqrt{t-x}} \ee (note that $\s(x)$ is not singular at $x=0$
because of the condition (15) below), so that we are able now to
calculate the total mass of the disk:
\be
M=2\pi\int\limits_0^R \s(\rho)\rho\,\d\rho= -\frac{R^3}{4\pi
G}\int\limits_0^1\frac{\d x}{x}\int\limits_x^1 \frac{\a(t)\,\d
t}{t\sqrt{t-x}}\,. \ee

For $M$ to be a finite quantity, it is necessary that
\be
\lim_{x\to0}\int\limits_x^1 \frac{\a(t)\,\d t}{t\sqrt{t-x}} =
\int\limits_0^1 \frac{\a(t)\,\d t}{t^{3/2}} = 0\,. \ee

From (14) and (15) a simpler formula for the mass of the disk is
readily obtainable:
\be
M=-\frac{R^3}{4\pi G}\int\limits_0^1\frac{\a(t)}{t^{3/2}}\ln t\,\d
t\,. \ee

Taking into account (11) and (15), the balance equation (10)
yields the integral equation of the Abel--type with respect to
$\a(t)$:
\be
-\frac{1}{2\pi} \int\limits_0^{x}\frac{\a(t)\,\d t}
{t^{3/2}\sqrt{x-t}}=\int\limits_0^{x}
(\omega^2t^{3/2})_{t}^{'}\frac{\d t}{\sqrt{t}}+A_0\,,\ee where the
integration constant $A_0$ should be found from the condition
(15).

The solution of equation (17) can be written in the form (cf.
Ref.~\cite{Sne}, p.~318)
\be
\frac{\a(x)}{2x^{3/2}}= -\frac{\d}{\d x}\int\limits_0^x \frac{\d
t}{\sqrt{x-t}}\Bigl( A_0+\int\limits_0^t
\frac{1}{\sqrt{s}}(\omega^2s^{3/2})_s^{'}\,\d s\Bigr) =
-\frac{A_0}{\sqrt{x}}
-\int\limits_0^x\frac{(\omega^2t^{3/2})_t^{'}\,\d t}
{\sqrt{t}\sqrt{x-t}}\,.\ee

In the general case the surface density has the following
asymptotic behavior in the neighborhood of the outer rim $x\to 1$:
\be
\s(x)\approx A\sqrt{1-x}\,, \quad A=\frac{2R}{\pi^2 G}
\int\limits_0^1\frac{(\omega^2t^{3/2})_t^{'}\sqrt{t}\,\d t}
{\sqrt{1-t}}\,.\ee

Formulas (13), (18) provide the {\it general} solution to the
problem under consideration in terms of two successive
quadratures: for a given distribution $\omega(\rho)$, one should
first find the function $\a(\rho)$ with the aid of the formula
(18) and then calculate the function $\s(\rho)$ from the integral
(13). The total mass of the disk is defined by (16).

Let us illustrate the use of the general formulas with some
concrete examples. Among numerous classes of smooth rotation
curves which can be treated analytically with the aid of our
formulas we choose the physically meaningful velocity
distributions of the type \be
\omega^2=\omega_0^2b^3/(b^2+\rho^2)^{3/2}, \quad \omega_0,b={\rm
const}. \ee

This particular curve was proposed by Brandt \cite{Bra} in the
case of {\it infinite} disks and it possesses the Keplerian
asymptotics at $\rho\to\infty$; its parameter $b$ multiplied by
$\sqrt{2}$ has the meaning of the radius of the circumference at
which the linear velocity has its maximal value, while $\omega_0$
is the angular velocity at the center of the disk. A wide family
of asymptotically Keplerian rotation curves having the form
\be
\omega_n^2=\frac{\omega_0^2\,b^{2n+2}}{n!}
\left(-\frac{\partial}{\partial b^2}\right)^{n-1}
\frac{1}{b(b^2+\rho^2)^{3/2}}\,, \quad n=1,2,... \ee and
containing (20) as the particular $n=1$ specialization was studied
by Toomre \cite{Too}; the corresponding surface--density
distributions (in the case of {\it infinite} disks) can be shown
to be given by the formula
\be
\s_n(\rho)=\frac{\omega_0^2\,b^{2n+2}(2n-1)!!}{2^nn!\,\pi
G(b^2+\rho^2)^{(2n+1)/2} } \,. \ee

In what follows we shall obtain the analog of the expression (22)
in the case of {\it finite} disks. For this purpose it is
advantageous to rewrite the rotation curves (21) in terms of the
dimensionless variable $x=\rho^2/R^2$ and parameter $a=b^2/R^2$:
\be
\omega_n^2=\frac{\omega_0^2\,a^{n+1}}{n!}
\left(-\frac{\partial}{\partial a}\right)^{n-1}
\frac{1}{\sqrt{a}(x+a)^{3/2}} \ee (all these $\omega_n$ take the
same value at $x=0$: $\omega_n(0)=\omega_0$).

Let us first consider the case $n=1$. Setting $n=1$ in (23) and
substituting the resulting $\omega_1^2$ into (18), we obtain
\be
\frac{\a_1(x)}{2x^{3/2}}=\omega_0^2\left(
\frac{\sqrt{x}(3a+x)a}{(a+x)^2}-\frac{2a}{1+a}\right)\,.\ee

The corresponding formula for the surface--density distribution is
then obtainable from (13), and it has the form
\be
\s_1(x)=\frac{\omega_0^2\,a^{3/2}b S}{\pi^2 G}\,, \quad
S\equiv\frac{\sqrt{(1-x)(a+x)}+(1+a){\rm
Arctan}\sqrt{(1-x)/(a+x)}} {(1+a)(a+x)^{3/2}}\,. \ee

Due to the linearity of the problem, the surface--density
distributions corresponding to the rotation curves (23) will be
defined in the general case by the expression
\be
\s_n(x)=\frac{\omega_0^2\,a^{n+\frac{1}{2}}R}{n!\,\pi^2 G}
\left(-\frac{\partial}{\partial a}\right)^{n-1} S\,. \ee

In the limit $R\to\infty$, formula (26) reduces to Toomre's
expression (22). In Fig.~1 we have plotted the surface--density
distribution (25) as the function of $\rho/b$ for different values
of the ratio $b^2/R^2$.

The total mass of the disk defined by the surface density (26) is
given by the formula
\be
M_n=\frac{2\omega_0^2\,a^{n+1}R^3}{n!\,\pi G}
\left(-\frac{\partial}{\partial a}\right)^{n-1} \left(
-\frac{1}{1+a}+\frac{1}{\sqrt{a}} {\rm
Arctan}\frac{1}{\sqrt{a}}\right)\,, \ee and it can be readily
found from (16).

It is worthwhile mentioning that the rotation curves can be also
approximated by linear combinations of different functions
$\omega_n^2$ defined by (21):
\be
\omega^2=\sum\limits_{n=1}^N c_n\omega_n^2\,, \quad
\sum\limits_{n=1}^N c_n=1\,, \ee and they will all possess the
limit of the rigid--body rotation with a given angular velocity
$\omega_0$ near the center and the Keplerian asymptotics at large
$\rho$. The corresponding surface--density distributions and total
masses will be given by the formulas
\be
\s(x)=\sum\limits_{n=1}^N c_n\s_n(x)\,, \quad
M=\sum\limits_{n=1}^N c_nM_n(x)\,, \ee and in the neighborhood of
the outer rim the surface density will have the following
asymptotics:
\be
\s(x)\approx\frac{\omega_0^2R\sqrt{1-x}}{\pi^2
G}\sum\limits_{n=1}^N c_n \left(\frac{a}{a+1}\right)^n\,. \ee

An interesting important question which we would like to discuss
in relation with our formalism is the possible estimation of upper
limits of the galactic masses and radii of outer rims since, as is
well known, real galactic disks can extend at much larger
distances than might be thought from the observations of hot gas
\cite{Sch}. Here we shall restrict our consideration to only the
case of analytic continuation of the rotation curves into the
region $\rho>R_v$ where $R_v$ is a known radius of the visible
part of the galactic disk.

Let an arbitrary smooth distribution $\omega^2(\rho)$ be known
from observational data for $0\le\rho\le R_v$. Then the
corresponding $\s(\rho)$ should be constructed via the formulas
(13) and (18) at the supposition that the real radius $R$ of the
disk is an unknown parameter to be determined. The upper limit
$R_{max}$ for the radius of the disk can be found from the
condition $(\d\s/\d\rho)_{\rho=R}=0$ or $A=0$ (see formula (19)),
and the value of $R_{max}$ thus obtained should be subsequently
substituted into the formula (16) to get the maximal value of the
mass. A different, though equivalent, way of obtaining $R_{max}$
is to find it from the condition of mass' maximum. Hence, for the
real radius $R$ and real mass $M$ of the galactic disk we get the
inequalities $R_v\le R\le R_{max}$ and $M\le M_{max}$. Mention,
that for some rotation curves $R_{max}$ may have finite values,
but it can also happen that $R_{max}=\infty$ (like in the case of
the pure Toomre's angular--velocity distributions $\omega_n^2$, so
that only a linear superposition of different $\omega_n^2$ leads
to finite $R_{max}$).

The existence of maximal values for disks' masses at finite radii
for the analytically continued rotation curves constitutes a new
physical phenomenon (very similar to the static criteria of
stability for the barotropic models of stars \cite{Tas}) which is
absent in infinite disks. For instance, in the case of the angular
velocity distribution (28) the condition $A=0$ yields the
algebraic equation
\be
\sum\limits_{n=1}^N c_n\left(\frac{a}{a+1}\right)^n=0\,, \ee As
can be easily seen, for all $c_n>0$, $n=\overline{1,N}$, this
equation has the unique non--negative root $a=0$ corresponding to
$R=\infty$. But already for $c_1<0$, $c_i>0$, $i=\overline{2,N}$,
it will have an additional, positive root $a=a_*>0$. In the
interval $0<a<a_*$ the surface density turns out to be negative in
the vicinity of the outer rim $x=1$, so expressions (29) are only
physically meaningful for $a\ge a_*$, the boundary value of the
disk radius being $R_{max}=b/\sqrt{a_*}$.

Let us illustrate the above said with the simplest linear
combination of the curves (20) possessing the Keplerian
asymptotics: \bea \omega^2&=&\frac{4q-1}{3}\omega_1^2+\frac{4}{3}
(1-q)\omega_2^2=\omega_0^2\left[ q\left(\frac{a}{a+x}\right)^{3/2}
+(1-q)\left(\frac{a}{a+x}\right)^{5/2}\right]\,, \nonumber\\ &&
0\le q\le 1, \quad 0<a<0.5\,. \eea

The corresponding surface--density distribution is given by the
formula
\be
\s(x)=\frac{4q-1}{3}\s_1(x)+\frac{4}{3}(1-q)\s_2(x)\,, \ee where
$\s_1$ and $\s_2$ are defined by (25).

The derivative $(\d\s/\d\rho)_{\rho=R}$ will be equal to zero when
$-1+3a+4q=0$, whence one obtains that $a$ is equal to $(1-4q)/3$.

On the other hand, the mass of the disk with the angular--velocity
distribution (32) is equal to
\be
M=\frac{2\omega_0^2b^3}{\pi}\left[\frac{\sqrt{a}(2-5q-3aq)}{3(1+a)^2}
+q{\rm Arctan}\frac{1}{\sqrt{a}}\right]\,, \ee and, as function of
$a$ for fixed $\omega_0^2b^3$, this expression achieves its
maximum at $a_*=(1-4q)/3$.

Therefore, we have arrived at the following important conclusion:

When $0\le q<0.25$, the radius of the disk cannot be greater than
$b\sqrt{3/(1-4q)}$. For $0.25\le q\le 1$, as can be seen from
(34), the mass of the disk becomes a monotonically decreasing
function of $a$, so that its maximum is achieved in the limit
$R\to\infty$. The dependence of $M_{max}$ on $q$ is shown in
Fig.~2. This example reveals in particular the difficulties which
one may encounter while approximating a real rotation curve by
superposing standard rotation curves in the case of infinite
disks; these difficulties are absent in the case of disks of
finite extension.


This work has been supported by Projects 34222--E and 38495--E
from CONACyT of Mexico. N.R.S. acknowledges financial support from
SNI--CONACyT, Mexico.

\begin{figure}[htb]
\centerline{\epsfysize=90mm\epsffile{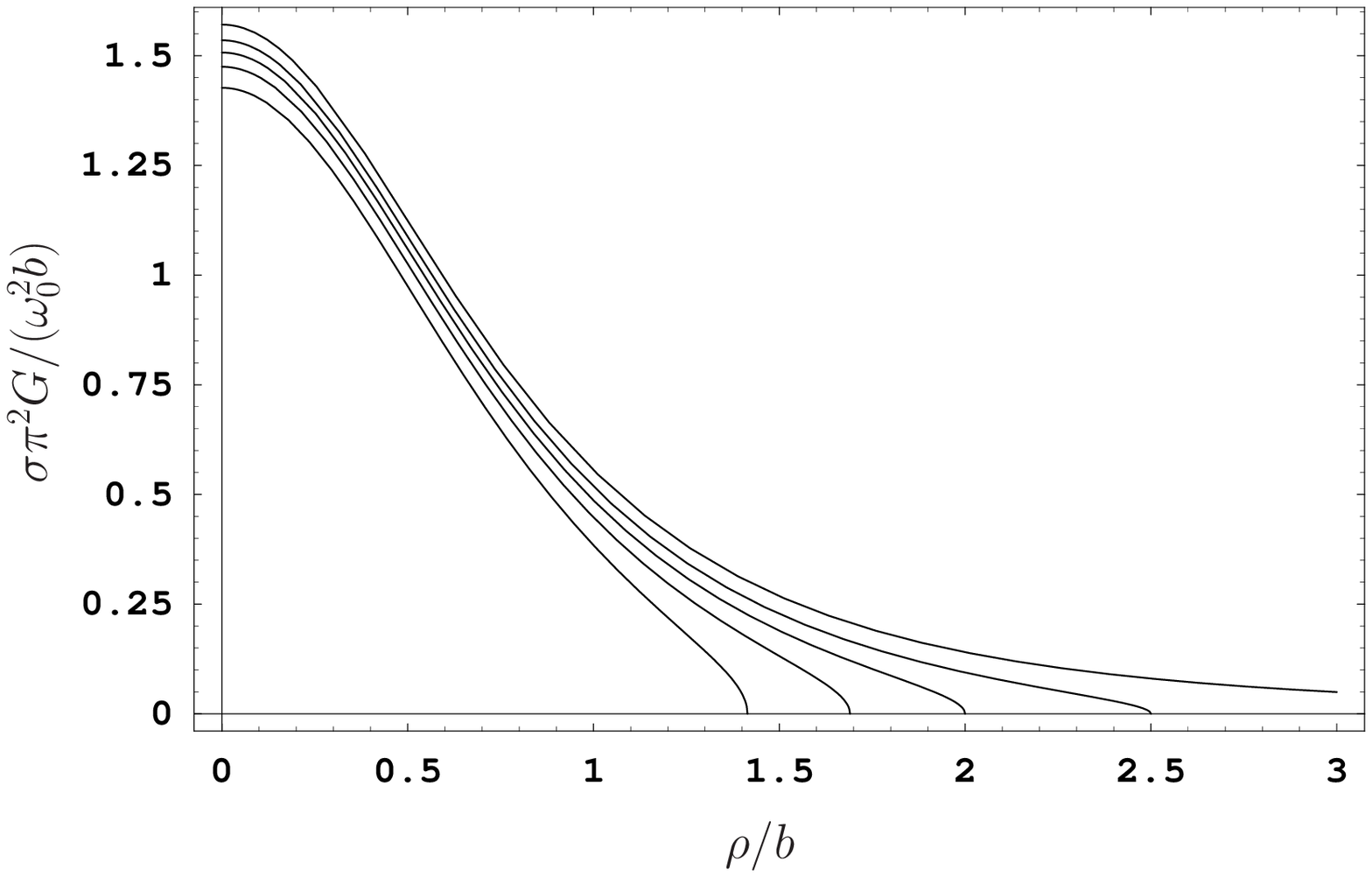}} \caption{The form
of the surface density distribution (25) for several particular
values of $a$ ($a=0,0.16,0.25,0.35,0.5$).}
\end{figure}

\begin{figure}[htb]
\centerline{\epsfysize=90mm\epsffile{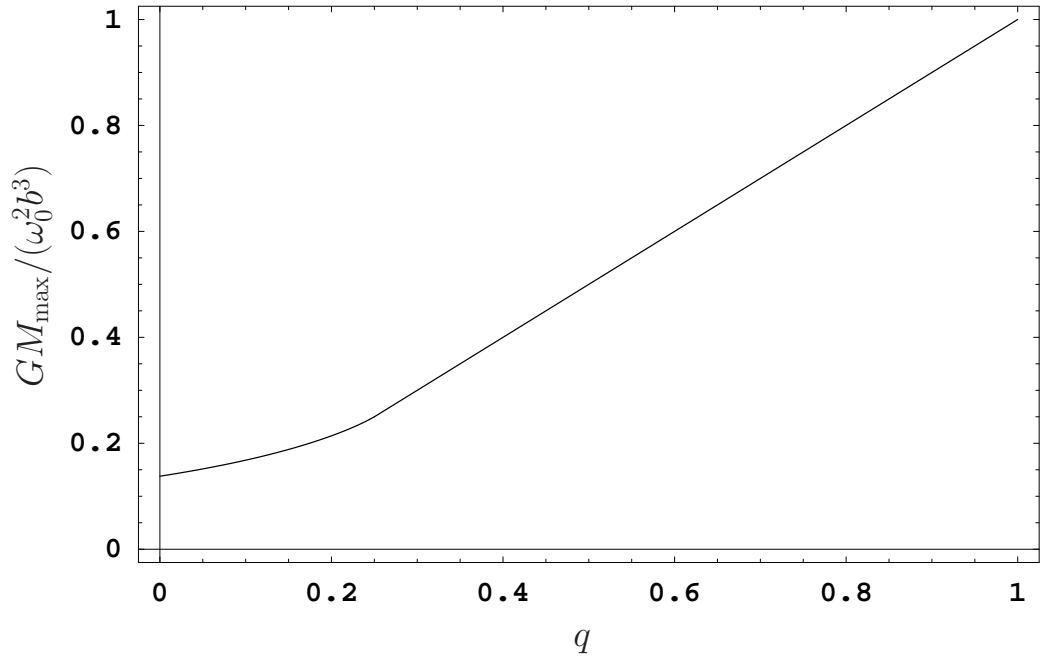}} \caption{The
dependence of the maximally possible mass of the disk on the
parameter $q$ in the example given by formulas (32)--(34).}
\end{figure}

\end{document}